\documentclass[journal]{IEEEtran}

\usepackage{amsmath, ,subfigure, dsfont }
\usepackage[final]{graphicx}
\usepackage{setspace}%
\usepackage{array}

\begin{document}

\title{Direct-Digital-Drive Microring Modulator}

\author{Yossef Ehrlichman,~\IEEEmembership{Member,~IEEE,}
        Ofer Amrani,~\IEEEmembership{Member,~IEEE}
        Shlomo Ruschin,~\IEEEmembership{Member,~OSA}

}

\maketitle

\begin{abstract}
The method of Direct Digital Drive is applied to a microring
resonator. The microring resonator is thus controlled by a segmented
set of electrodes each of which is driven by binary (digital)
signal. Digital linearization is performed with the aid of digital
memory lookup table. The method is applied to a single microring
modulator to provide an M-bit digital-to-analog converter (DAC),
which may also be viewed as an M-level pulse amplitude modulator
(M-PAM). It is shown, by means of simulation, that a 4-bit DAC can
achieve an effective number of bits (ENOB) of 3.74bits. Applying the
same method for two rings, enables the generation of two-dimensional
optical M-QAM signals. It is shown, by means of simulation, that a
16-QAM modulator achieves an EVM better than -30dB.
\end{abstract}


%
\section{Introduction}
%
%
\IEEEPARstart{A}{ttempting} to apply photonic technologies to data
processing has been taking place for several decades. Yet, their
implementation in practical systems remained limited and failed to
penetrate practical processors. Many demonstrated optical devices
are still 
oversized, were based on exotic materials and
needed rather complicated interfacing with other electronic
components. This paper takes advantage of recent developments in
Silicon microphotonics and novel interfacing schemes in order to
realize photonic components for fast communication and signal
processing. In addition to its practical potential, as presented,
the ideas exposed herein present basic physical and mathematical
challenges.
\\
Microring modulators were  proposed and demonstrated for analog
signal modulation and for simple digital modulation, such as
On-Off-Keying (OOK). Recently, microring modulators were
investigated for advanced digital modulation formats such as
PAM~\cite{hai2014low}~\cite{ding2014optical}~\cite{karimelahi2015pam},
QPSK~\cite{dong2013silicon} and even
QAM~\cite{ehrlichman2013generating}~\cite{karimelahi2016quadrature}.
\\
All the works mentioned above (\cite{karimelahi2016quadrature}
excluded) use an analog voltage signal to drive the modulator. The
application of analog signals usually calls for mediating electronic
circuitry, such as digital-to-analog conversion.
In the current work we promote 
the application of the, so-called, \textit{Direct Digital
Drive}~(DDD) ~\cite{Ehrlichman2008} method for use with microring
resonators.
DDD allows the utilization of only two voltage levels directly on
the photonic device; it makes the need for mediating devices, such
as electrical digital-to-analog converter, unnecessary.
\\
In the first part of this work 
we present a design of an N-bit digital-to-analog converter. A
digital-to-analog converter is a device that converts an N-bit
digital word to a corresponding analog (voltage) representation.
A 4-bit DAC produces 16, equally spaced analog levels
and can therefore be viewed as (a digitally controlled) 16-PAM
modulator. 
In the second part of this work 
we extended a previous work~\cite{ehrlichman2013generating}
for generating M-QAM signals with microring modulators
by utilizing the DDD approach. 
An all-digital M-QAM modulator is thus presented.

%
\section{Multi-electrode microring Digital-to-analog converter/M-PAM modulator}
%
%

\begin{figure}[ht] \centering
\includegraphics[width=8cm]{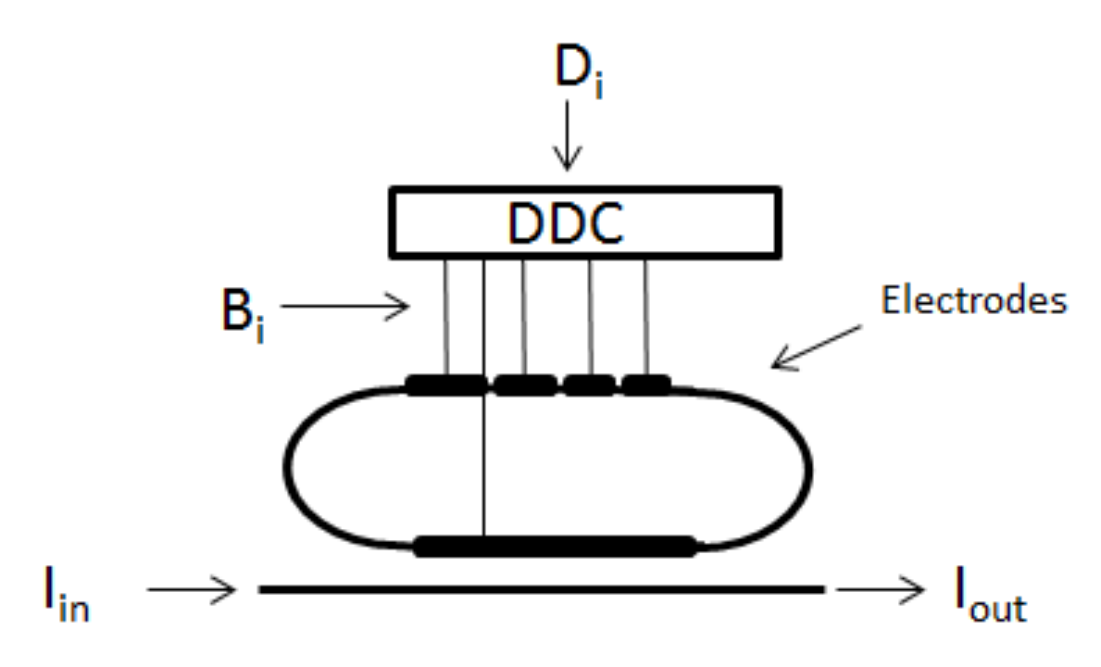}
\caption{Schematic illustation of a multi-electrode microring
DA.}\label{fig:mring_da}
\end{figure}

As an example, consider 
the 4-bit optical DAC 
based on a microring modulator, illustrated
in Figure~\ref{fig:mring_da}. The  device basic layout is similar to
previously published microring resonators. A CW light of wavelength
$\lambda$ and intensity $I_{in}$ is coupled from the waveguide to
the microring.
The coupling coefficient between the waveguide and the microring is
denoted by $t$ and the loss per round trip inside the microring is
$a$.
The light inside the microring is 
modulated by several, in this example $M=5$, independent phase
shifter segments (electrodes). The phase shifter can be implemented
as reversed-bias pn-diode or as zig-zag pn diode, etc.
Hereinafter, each phase shifter will be referred to as an \textit{electrode}.

At its electrical input, the device 
accepts an $N=4$ bits digital 
word, denoted  $D_i$, where $i=\{1\dots 2^N\}$. The input word is
mapped onto each of the $5$ electrodes via the digital-to-digital
converter (DDC). In essence, the DDC converts a $4$-bit input to a
$5$-bit output, which, in turn, 
control the $5$ electrodes. Note that each electrode is driven by
one of two voltage levels, $0$ and $V_1$, representing
binary $0$ and $1$.
Described as such, the DDC is basically a lookup-table that can be realized by 
a (high speed) digital memory. 
%

An optical modulator is typically characterized by the product
$V_{\pi}\cdot L$ where $V_{\pi}$ is the voltage 
whose application to an electrode of length $L$, 
induces a phase shift of $\pi$.
Without loss of generality, we set $L=2\pi R_1$, i.e. as the
circumference of the microring, where $R_1$ is the radius of the
ring.
Let $j$,  denote the index 
of the electrodes 
(in this example $j=1\dots 5$). Assume that the length of each
electrode is given by: $L_j=L\cdot 2^{-j}$. Note that $L > \sum
L_j$.

If a voltage $V_1$ is applied to electrode $j$, the induced phase
shift $\phi_j$ will be:
\begin{equation}
\phi_j=k\frac{2\pi}{\lambda}\Delta nb_{j}L_j,
%
%
\end{equation}
where $\Delta n$ is the effective refractive index modulation due to
the applied voltage $V_1$ on phase shifter $j$, $k$ is an empirical
constant that accounts for both the optical confinement and the
coefficient of the charge density induced refractive index change.
%
%
The parameter $b_j$ is a binary quantity that indicates whether
voltage $V_1$ was applied to phase shifter $j$. The dependance
between $\Delta n$ and the applied voltage $V_1$ is known to be a
nonlinear function, $f(V_1)$. The exact relation depends on the
phase shifter design. The intensity transmission of the DDD
multi-electrode microring structure can be written as:
\begin{equation}
\label{eq:dac_mring_transfer_red}
I_{out}=I_{in}\frac{a^2-2ta\cdot
cos\left(\sum_{j=1}^Mb_{ij}L_j\right)+t_1^2}{1-2ta\cdot
cos\left(\sum_{j=1}^Mb_{ij}L_j\right)+t^2a^2},
\end{equation}
where a binary matrix $B_i=\{b_{ij}\}$ of dimensions $M\times N$
holds the mapping of the $N=4$-bit input word $D_i$ on the $M=5$
electrodes - evidently, a highly non-linear transmission.

%

%
\subsection{M-PAM: the design process}
%
%
\label{sec:MPAMDesign}
%
To maximize the output 
dynamic range (DR), the microring resonator should be in critical
coupling, $a=t$. This will allow the smallest 
possible output intensity level to approach $0$.

\begin{figure}[h] \centering
\includegraphics[width=7cm]{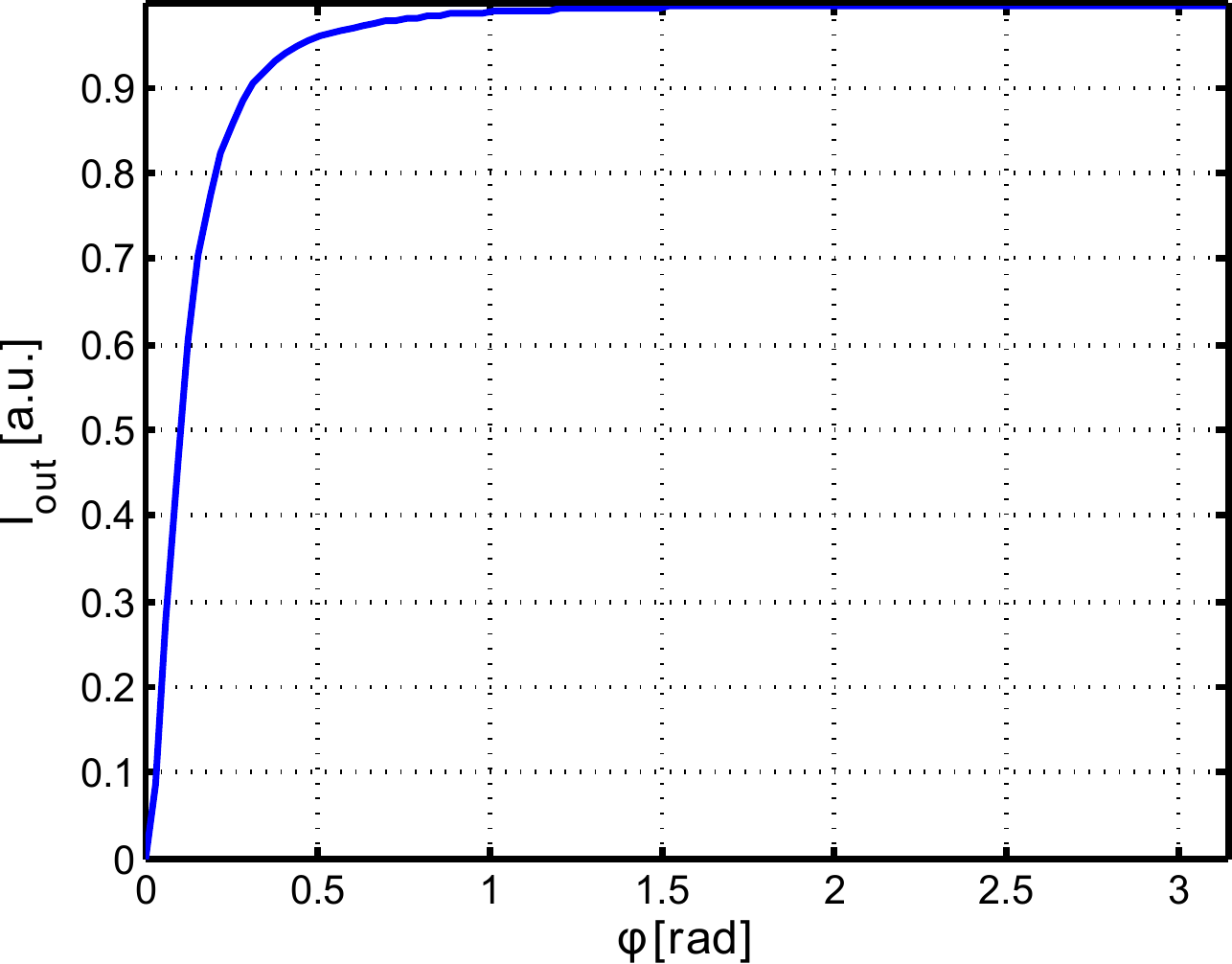}
\caption{The output intensity of a microring resonator for phase
shifts between $0$ and $\pi$.}\label{fig:ringresoIout}
\end{figure}

Figure~\ref{fig:ringresoIout} shows the output intensity of a
microring resonator for phase shifts between $0$ and $\pi$. Phase
shifts greater 
than about $0.5[rad]$ contribute very little
to the output dynamic range because of the
high nonlinearity
of the microring modulator. Hence, the voltage $V_1$ will be set so
as to
produce 
the intensity 
$I_{\phi_1}=I_{in}-\Delta I$.
%
%
%
%
The quantity $\Delta I$ is chosen as a tradeoff between (best
achievable) 
linearity and output dynamic range. 
Smaller values of $\Delta I$ can 
increase the output dynamic
range, but also will reduce the linearity of the device, unless
additional electrodes are added. A dynamic range of about $90\%$ of
the total available DR will keep the electrode count low and close
to the input bit length.
In this example 
$V_1$ was set to induce a phase shift
that leads to an output dynamic range of about $93\%$.

\begin{figure}[ht] \centering
\includegraphics[width=7cm]{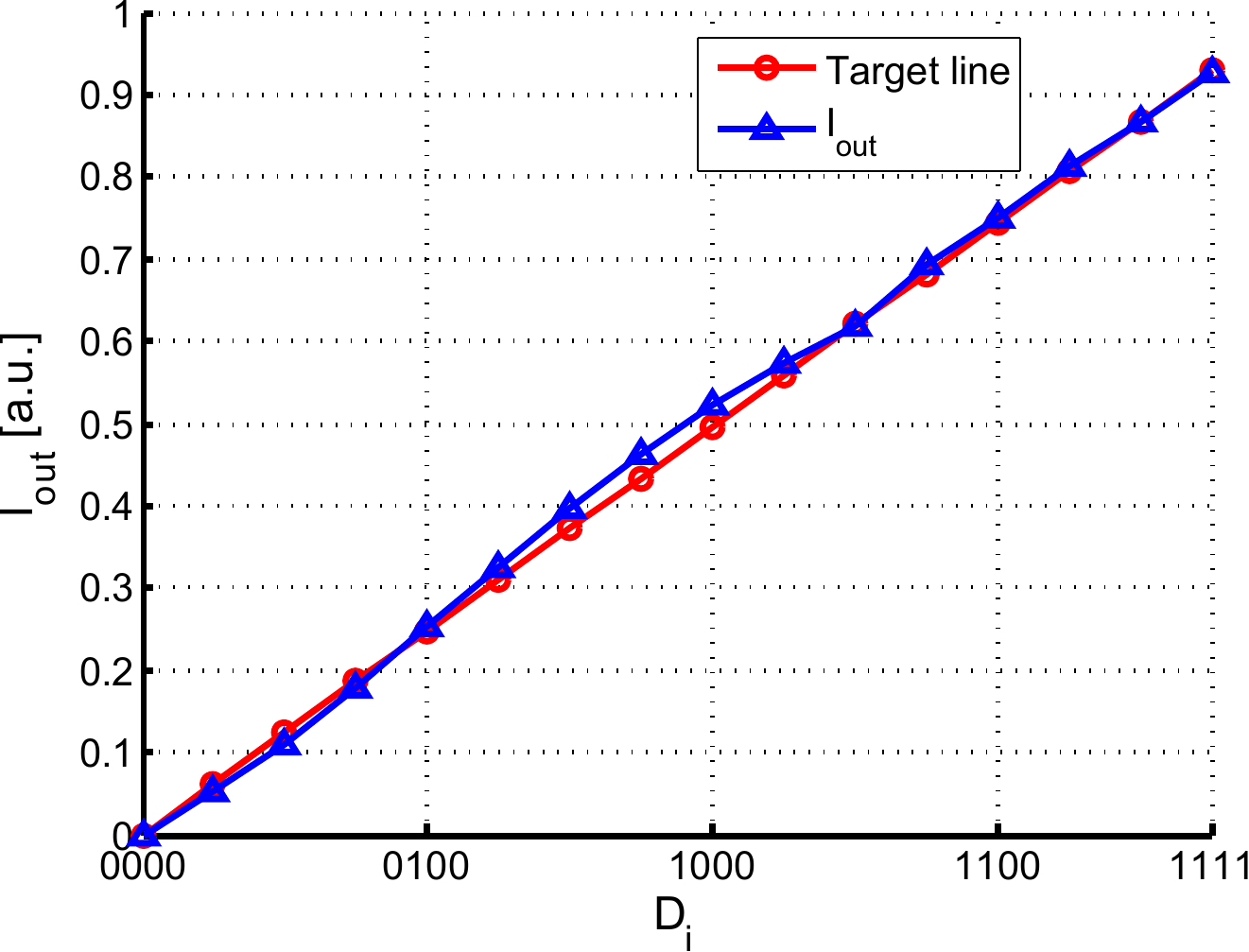}
\caption{The output intensity of an ideal output of a 4-bit DAC (16
levels on a straight line) compared to a DDD microring modulator
with 5 electrodes.}\label{fig:IoutN=4M=5}
\end{figure}

\begin{figure}[ht] \centering
\includegraphics[width=7cm]{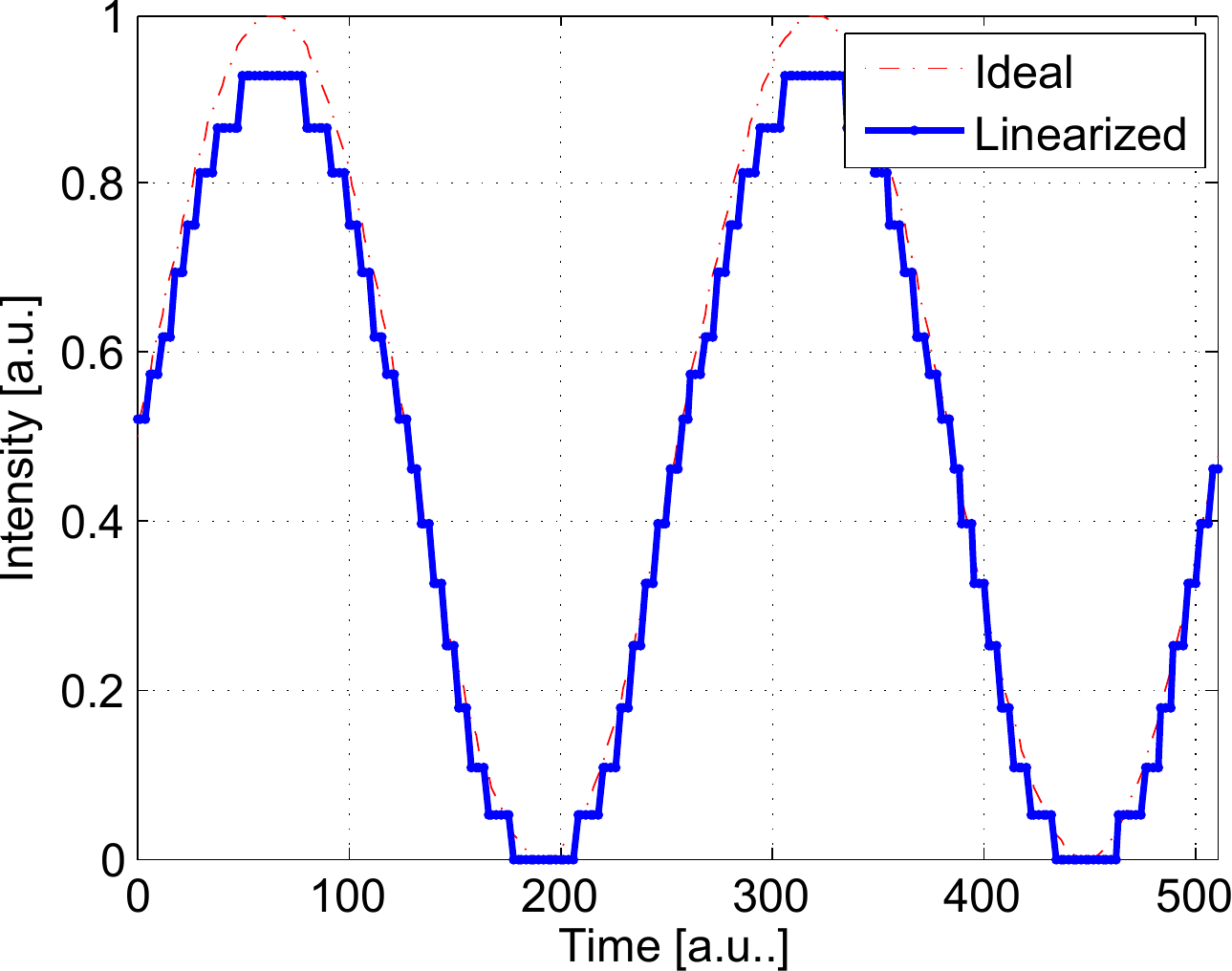}
%
\caption{A sinewave generated by the 4-bit DAC of Fig.
\ref{fig:mring_da} \label{fig:sinewaveN4M5}}
\end{figure}

Figure~\ref{fig:IoutN=4M=5} shows the output intensity of
a 4-bit DAC (16 levels on a straight line), based on a microring
modulator with 5 electrodes.
%
%
Figure~\ref{fig:sinewaveN4M5} shows a sinewave generated with the
proposed device. The linearity of the DAC can be quantified by
standard figure of merits: its Differential Non-Linearity (DNL) is
0.2 bits; the Integral Non-Linearity (INL) is 0.4bits and the
Effective Number of Bits (ENOB) is 3.74bits.

%

%
\section{Multi-electrode microring M-QAM modulator}
%
%
\label{sec:QAM_modulator_construction}
In the previous sections we presented the design of a DAC, which is
equivalent to a PAM modulator, by using a single microring
modulator with several electrodes driven by a digital signal.
%
In this section we use a  
two microring configuration 
to generate M-QAM (two-dimensional) signal constellations.
Quadrature amplitude modulation (QAM) is a modulation scheme that
conveys data by means of modulating both the amplitude and the phase
of a sinusoid carrier thus providing spectral efficiencies in excess
of 2bit/symbol.
In 
previous work~\cite{ehrlichman2013generating}, we introduced a
method for generating optical M-QAM signals by using two mutually
decoupled microring modulators.
In this configuration, one ring was used to generate the amplitude
of the desired signal (perturbed by some deterministic phase
shift), while the second ring was used to complement 
the phase required to obtain
the desired complex signal.
The electronic input to this modulator was an analog voltage.

Herein, in order to drive the two rings with digital (two-level) signals, we split 
the electrode into segments. 
An M-QAM modulator is schematically depicted in
Figure~\ref{fig:mring_qam}.

%
\subsection{M-QAM: the design process}
%
%
The design process is similar to the one described in Section
\ref{sec:MPAMDesign}
except that we now have 
to generate complex signals 
rather than intensity levels only.
The role of the 
modulator is to generate a specific 
constellation of points consisting of 
$M$ distinct complex points (also refereed to as {\it signals}); the
constellation points can be generally formulated as follows:
\begin{equation}
\label{eq:target_signals}
s_i=r_ie^{j\theta_i},\quad r_i>0,\quad 0\leq\theta_i\leq 2\pi,\quad
i=1,\dots,2^M.
\end{equation}

\begin{figure}[!ht]
   \centering
        \includegraphics[width=8cm]{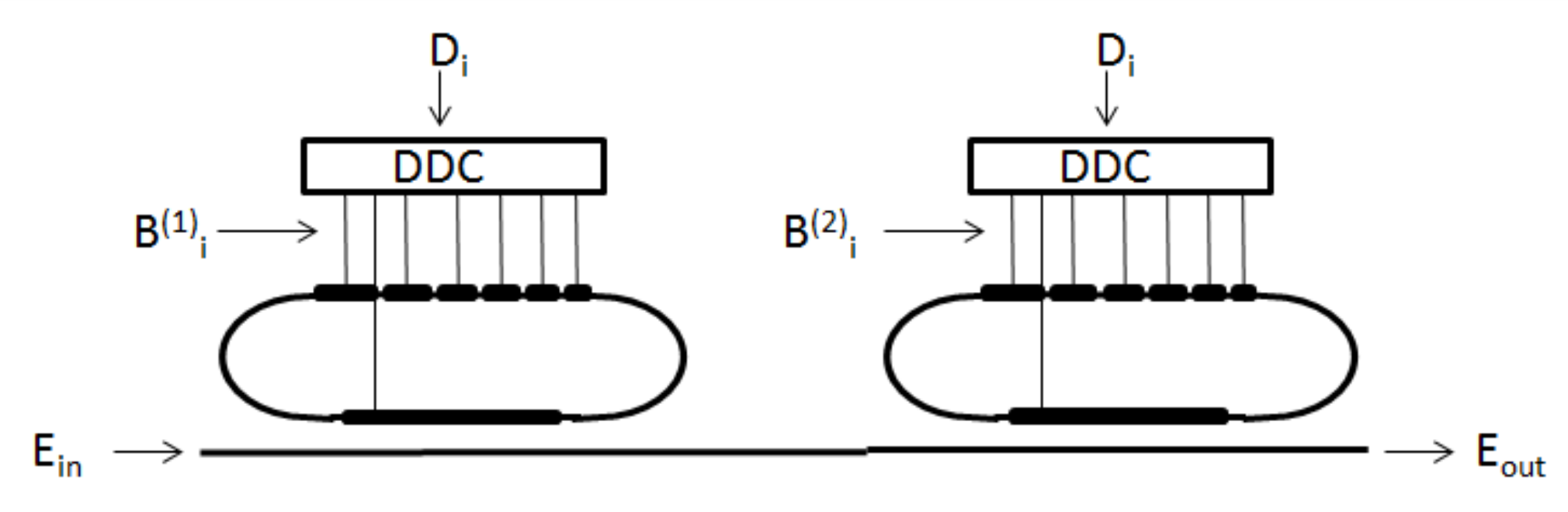}
    \caption{16QAM Modulator based on two, multi-electrode, DDD micro-ring resonators.
    A 4-bit digital input, $\mathbf{D_i}$, is mapped by the DDC to a 14-bit
    output that drives the 14 segmented electrodes.
The electrode lengths (per micro-ring) follow a divide-by-two
sequence. The DDC, shown here for exposition purposes as two
separate boxes, is actually a single memory device.
}
    \label{fig:mring_qam}
\end{figure}

The proposed modulator is described by means of an example.
Figure~\ref{fig:mring_qam} depicts a 16QAM optical modulator based
on two micro-ring resonators equipped with multiple electrodes.  The
electrodes in this example are divided between the two micro-rings:
$7$ electrodes on each micro-ring.
%

As input, this QAM modulator accepts a 4-bit digital 
word, denoted $\mathbf{D_i}$.
The 4-bit input word is mapped onto each of the $14$ electrodes via
the DDC. 
%
%
%
Thus, each electrode is driven by one of two voltage levels, $0$ or
$v$, representing binary $0$ and $1$, respectively.

More generally, Let
$\mathbf{L^{(1)}}=(L^{(1)}_1,L^{(1)}_2,...L^{(1)}_{N_1})$ and
$\mathbf{L^{(2)}}=(L^{(2)}_1,L^{(2)}_2,...L^{(2)}_{N_2})$ be two
vectors of dimensions $N_1$ and $N_2$, whose elements correspond to
the lengths of the electrodes of the first microring and the second microring, respectively.
Let $\mathbf{B}^{(1)}$ be a binary $M\times N_1$ matrix. Row $i$ of
$\mathbf{B}^{(1)}$, ${B}^{(1)}_i$, holds the mapping from input
$D_i$ onto the $N_1$ electrodes of the left ring. 
Likewise, binary matrix $\mathbf{B}^{(2)}$, of dimensions $M\times
N_2$, holds the mappings from each of the input digital words to the
$N_2$ electrodes of the right ring.
With this nomenclature, the output of the modulator can be
formulated as a function of the~digital~input~$D_i$:
\begin{equation}
\label{eq:mring_qam_output}
E_{out}(D_i)=E_{in}exp\left[j\left(\pi+\phi_1\right)\right]\frac{\alpha_1-t_1 exp{-j\phi_1}}{1-t_1\alpha_1 exp\left(j\phi_1\right)}\cdot
\end{equation}

\begin{equation*}
\label{eq:mring_qam_output_cont}
\cdot exp\left[j\left(\pi+\phi_2\right)\right]\frac{\alpha_2-t_2 exp{-j\phi_2}}{1-t_2\alpha_2 exp\left(j\phi_2\right)}
\end{equation*}
and the phase shift are

\begin{equation}
\label{eq:mring_qam__phases}
\phi_1=\pi\frac{v}{v_{\pi}}\sum^{N_1}_{j=1}B^{(1)}_{ij}L^{(1)}_j
\end{equation}
\begin{equation}
\label{eq:mring_qam__phases_cont}
\phi_2=\pi\frac{v}{v_{\pi}}\sum^{N_2}_{j=1}B^{(2)}_{ij}L^{(2)}_j
\end{equation}
where $E_{in}$ denotes the amplitude of the optical field entering
the modulator.

The geometrical structure and the loss and coupling parameters are set in a similar manner to
the description in~\cite{ehrlichman2013generating}.
The first ring is designed to work in critical coupling for 
which $t=a$ in order to generate the largest span of amplitudes. The second microring is designed to work in under-coupling regime, $t<a$, to act as a phase shifter with minimum amplitude loss.
%
By applying all possible combinations of the digital words $B_i$, 
a finite pool of points, of maximum cardinality $2^{N_1+N_2}$, can be generated.

\begin{figure}[!ht]
   \centering
        \includegraphics[width=8cm]{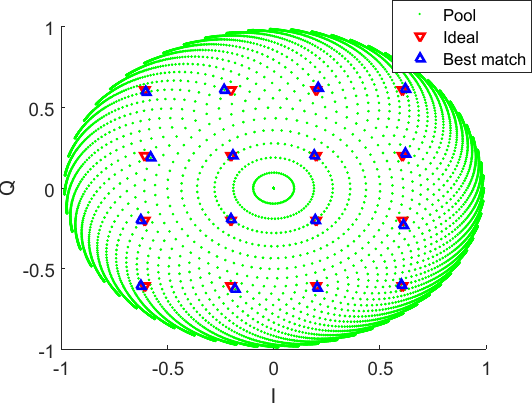}
    \caption{16QAM with $N_1=N_2=7$. 
    The EVM obtained is $-30.6dB$.}
    \label{fig:16QAM_N12=7}
\end{figure}

As an example, a 
pool of (green) points is shown 
in Figure~\ref{fig:16QAM_N12=7} for $N_1=N_2=7$.
From this large pool of possible points, one can choose a finite set of points that form a constellation for digital optical communications, $16$-QAM in this example.
%
An ideal set of points that constitute a $16$-QAM constellation,
as suggested by Eq.~\ref{eq:target_signals} with $M=4$,
is portrayed by the red triangles in Figure~\ref{fig:16QAM_N12=7}.
Out of the green pool of $2^{7+7}$ signal points,
we first select
the $16$ signals that provide the best match for the ideal $16$ points.
The selected set of best points amounts to a mapping between $16$ $D_i$ digital input values and the corresponding $B_i$'s, and is executed by the DDC.
%

As 
can be seen in the figure, the selected  points are not identical to the ideal points, thus producing an error.
%
To quantify this error, we employ the Error Vector Magnitude (EVM)~\cite{ehrlichman2013generating} measure. For the above example the EVM is $-30.6dB$.
Note that the EVM can be further reduced by increasing the number of electrodes in each microring. Such an increase will allow generating a denser pool of points from which one can select a desired set of constellation points with higher accuracy. 
Table~\ref{tbl:QAM_EVM} presents 
the obtained EVM for various constellations and varying number of electrodes.
%
It shows the EVM for 16-QAM with $6+6$ and $7+7$ electrodes. The difference between these 
configurations is $9$dB.
It can be seen that the inherent nonlineariy of the microring leads to a more involved 
 implementation of M-QAM modulators compared to M-PAM modulators.
The table also presents results for 64-QAM and 256QAM. While 
for 64QAM we achieve
EVM better than -30dB with $7+7$ electrode configuration,  256QAM requires 
one additional electrode 
in the amplitude-related modulator ring to achieve such EVM.
\\
\begin{table}[!t]
\renewcommand{\arraystretch}{1.3}
\caption{ EVM as a function of number of electrode}
\label{tbl:QAM_EVM}
\centering
\begin{tabular}{|c|c|c|c|}
\hline
$M$-QAM & $N_1$ & $N_2$ & EVM [dB]\\
\hline
16&6&6&-21.8\\
16&7&7&-30.6\\
64&7&7&-30.4\\
256&7&7&-26.9\\
256&8&7&-30.9\\
\hline
\end{tabular}
\end{table}
%

%
\subsection{Higher order QAM constellations}
\begin{figure}[!ht]
   \centering
        \includegraphics[width=8cm]{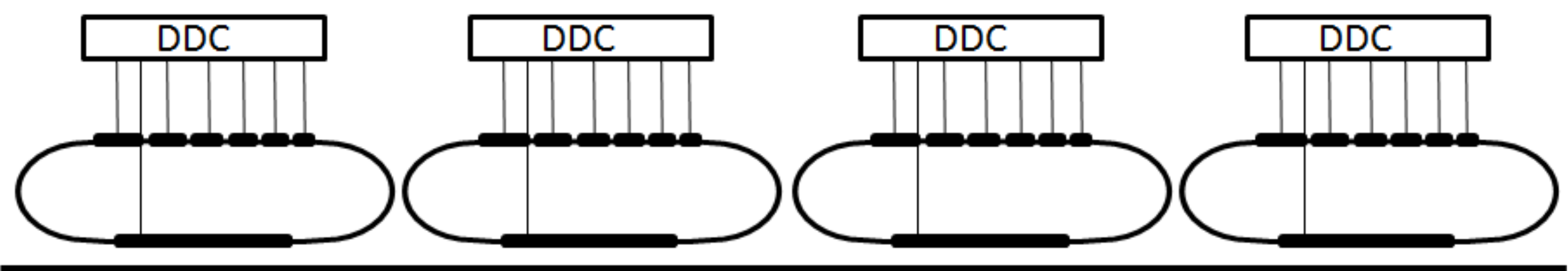}
    \caption{Multi micron configuration. The number of microring has been increased, but the number of electrodes per microring was reduced.}
    \label{fig: microring_qam_P=2}
\end{figure}

%
%
The number of  
distinct electrode segments that can be %
effectively mounted 
on a single microring is obviously limited. This, however, does not limit the proposed technology,
as the original configuration can be augmented by additional rings with additional electrodes. 
Thus for example, rather than generating $256$-QAM with $7$ and $8$ electrodes on each ring, one can utilize 
$4$ rings with $5$ electrodes on each ring: two of the rings will act as amplitude modulators while the other two will act as phase modulators, as depicted in Figure~\ref{fig: microring_qam_P=2}. For the same parameters as above, with the new configuration of $4$ microrings, EVM of $-33dB$ is achieved.
%

%
\section{Practical Implementation}
%
%
The practical implementation of the DDD modulators is not obvious. Next, we briefly 
discuss issues that are associated with the implementation 
of the above devices in 
silicon photonic technology.
Silicon has 
the potential of 
full integration of optics and electronics either in a monolithic or a hyrbrid process.
The first problem that arises with silicon is that the plasma dispersion effect (which is used widely to make phase shifters for silicon modulators) induces a voltage dependent loss. Meaning, it is impossible to make a lossless phase shifter. However, this problem can be alleviated 
by the DDD approach.
More specifically, the problem is that the VDL (Voltage Dependent Loss) will reduce the average optical power and hence "compress" the signal pool. The solution is simply to configure the DDC
to map (shrink) 
the required constellation to a lower average optical power.

The second problem that might arise is that in order to generate a phase shifter to induce phases between $0$ and $2\pi$, a large voltage signal will be needed. This problem can be solved by using the multiple microring configuration as discussed in the previous section. For example, instead of using a single ring to induce the $2\pi$ phase shift, we can use multiple modulators to induce the phase we are targeting.

%
\section{Conclusion}
%
%
We presented application of the Direct Digital Drive approach to microring resonators. We showed that this approach enables one to realize 
a digitally driven optical devices, that include  
compact M-PAM modulator and DACs. Extending the approach to more then one microring, enables the generation of optical M-QAM constellations. We showed that it is possible to generate any constellation order 
either by using 
a large number of electrodes, or by using a small number of electrodes 
but employ additional microring modulators.

%

\end{document}